\documentclass[a4paper,10pt,eqno]{article}
\usepackage{theorem}
\usepackage{latexsym,amssymb,amsfonts,amsmath}
\usepackage{cite}
\usepackage{color}
\usepackage{comment}

\newtheorem{thm}{Theorem}[section]
\newtheorem{lem}[thm]{Lemma}
\newtheorem{cor}[thm]{Corollary}
\newtheorem{pro}[thm]{Proposition}
\newtheorem{rem}[thm]{Remark}
\newtheorem{ass}[thm]{Assumption}

\theoremheaderfont{\scshape}
\newcommand{\RM}{\mathbb{R}}

\newcommand{\CM}{\mathbb{C}}

\newcommand{\PM}{\mathbb{P}}

\newcommand{\qed}{\hfill $\Box$}

\newcommand{\ket}[1]{|#1\rangle}
\newcommand{\bra}[1]{\langle#1|}

\title{{\Large {\bf A spectral analysis of discrete-time quantum walks related to the birth and death chains}}}
\author{
{\small Choon-Lin Ho}\\
{\scriptsize Department of Physics, 
Tamkang University}\\
{\scriptsize Tamsui 251, Taiwan (R.O.C.)}\\
{\scriptsize e-mail: hcl@mail.tku.edu.tw}\\
\\
{\small Yusuke Ide}
\footnote{To whom correspondence should be addressed. E-mail: ide@kanagawa-u.ac.jp}\\
{\scriptsize Department of Information Systems Creation, 
Faculty of Engineering, 
Kanagawa University}\\
{\scriptsize Kanagawa, Yokohama 221-8686, Japan}\\
{\scriptsize e-mail: ide@kanagawa-u.ac.jp}\\
\\
{\small Norio Konno}\\
{\scriptsize Department of Applied Mathematics, 
Faculty of Engineering, 
Yokohama National University}\\
{\scriptsize Hodogaya, Yokohama 240-8501, Japan}\\
{\scriptsize e-mail: konno@ynu.ac.jp}\\
\\
{\small Etsuo Segawa}\\
{\scriptsize Graduate School of Information Science, 
Tohoku University}\\
{\scriptsize Aoba, Sendai 980-8579, Japan}\\
{\scriptsize e-mail: e-segawa@m.tohoku.ac.jp}\\
\\
{\small Kentaro Takumi}\\
{\scriptsize Department of Applied Mathematics, 
Faculty of Engineering, 
Yokohama National University}\\
{\scriptsize Hodogaya, Yokohama 240-8501, Japan}\\
{\scriptsize e-mail: tororo224@yahoo.co.jp}\\
}
\date{\today }
\begin{document}
\maketitle

\par\noindent
\begin{small}
\par\noindent
{\bf Abstract}
\newline 
In this paper, we consider a spectral analysis of discrete time quantum walks on the path. For isospectral coin cases, we show that the time averaged distribution and stationary distributions of the quantum walks are described by the pair of eigenvalues of the coins as well as the eigenvalues and eigenvectors of the corresponding random walks which are usually referred as the birth and death chains. As an example of the results, we derive the time averaged distribution of so-called Szegedy's walk which is related to the Ehrenfest model. It is represented by Krawtchouk polynomials which is the eigenvectors of the model and includes the arcsine law.
%\footnote[0]{
%{\it Abbr. title:} Spectral analysis of DTQWs on the path
%}
%\footnote[0]{
%{\it AMS 2000 subject classifications: }
%60F05, 60G50, 82B41, 81Q99
%}
%\footnote[0]{
%{\it PACS: } 
%03.67.Lx, 05.40.Fb, 02.50.Cw
%}
\footnote[0]{
{\it Keywords: } 
Quantum walk, Birth and death chain, Ehrenfest model, Krawtchouk polynomials
}
\end{small}

\setcounter{equation}{0}
%%%%%%%%%%%%%%%%%%%%%%%%%%%%%%%%%%%%%%%%%%%%%%%%%%%%%%%
\section{Introduction}
%%%%%%%%%%%%%%%%%%%%%%%%%%%%%%%%%%%%%%%%%%%%%%%%%%%%%%%
During the last two decades, the study of quantum walk has been extensively developed in various fields. This wide range developments are found in review articles such as Kempe \cite{Kempe2003}, Kendon \cite{Kendon2007}, Venegas-Andraca \cite{VAndraca2008, VAndraca2012}, Konno \cite{Konno2008b}, Manouchehri and Wang \cite{ManouchehriWang2013}, and Portugal \cite{Portugal2013}. From the mathematical point of view, discrete time quantum walks (DTQWs) are viewed as a quantum counterpart of discrete time random walks (DTRWs). Since DTRWs are very simple models, so they play fundamental and important roles in both theoretical fields and applications. Thus it is expected that DTQWs also play fundamental and important roles in various fields. There is a variant of DTQW so-called Szegedy's walk which is directly related to the DTRW \cite{Szegedy2004}. There are papers \cite{IdeKonnoSegawa2012, Segawa2013, IdeEtAl2014, PortugalSegawa2017, BaluLiuVAndraca2017} to reveal spectral properties and the time averaged probability of Szegedy's walk. One of the main objective of the study of the DTQWs is to make it clear the probability distribution of the walker. But in many cases, it is difficult to obtain rigorous expression of the distribution. In such cases, we study the time averaged probability and the stationary measures at first to understand averaged behavior of the walker. For example, in cycle graph cases, the probability distributions of DTQWs do not converge but the time averaged probabilities do converge to their limit distributions \cite{AharonovEtAl2001, BednarskaEtAl2003}.

In this paper, we focus on DTQWs on the path graph. At first, we make a connection between DTQWs and the corresponding birth and death chains. This is nothing but the inverse problem for Szegedy's walk cases. The correspondence between DTQW and the DTRW (birth and death chain) is simple but the equivalence of its spectrum of the two Jacobi matrices (Lemma \ref{lem:JacobiQWandRW}) is not trivial. For example, the equivalence is broken in the cycle graph cases\cite{AraiEtAl2016}. Furthermore, we develop a procedure for building eigenvectors of the Jacobi matrix of DTQWs from that of the transition matrix with reversible measures of the corresponding DTRW (Proposition \ref{pro:SpecDTRWandDTQW}). This type of direct correspondence between DTQWs and DTRWs has not known.

Next we consider a spectral decomposition of the time evolution operator of DTQWs with isospectral coins. Note that under Assumption \ref{ass:coinQW}, all the coins are isospectral but need not be the same because it allows different eigenvectors. We can construct the eigenvalues and eigenvectors of the time evolution operator of the DTQWs from that of the Jacobi matrix (Lemma \ref{lem:eigenUformJ}). Theorem \ref{thm:Ave} and its corollary (Corollary \ref{cor:Ave}) are the main result of this paper. This theorem shows that under Assumption \ref{ass:coinQW}, the time averaged distribution of the DTQW is described by the pair of eigenvalues of the coins as well as the eigenvalues, eigenvectors and the stationary distribution of the corresponding birth and death chain. Calculating the time averaged distributions of DTQWs corresponding to various birth and death chains to reveal the common properties of the DTQWs can be an interesting future problem. 

As an example, we derive the time averaged distribution Eq. \eqref{eq:aveEhrenfest} of Szegedy's walk related to the Ehrenfest model which has been considered in an analysis of DTQW on the hypercube \cite{MarquezinoEtAl2008}. It is represented by Krawtchouk polynomials which is the eigenvectors of the model and includes the discrete version of arcsine law. Making the scaling limit of this model clear like \cite{IdeKonnoSegawa2012} can be an interesting future problem. We can also consider cases related to various urn problems including the P\'olya's urn. In the general urn cases, the treatment of the model is more difficult because the total number of the balls changes in time. Therefore we only discuss the Ehrenfest model which preserves the total number of balls.

This paper organized as follows. In Sect. \ref{defDTQW}, we define DTQWs on the path graph and the time averaged distribution of it. We construct the corresponding DTRW (birth and death chain) of DTQW in Sect. \ref{connectionQWandRW} by using spectral information for DTQW and DTRW. Section \ref{spectralDTQW} is devoted to the proposed spectral analysis for DTQWs. The main results of this paper are stated in Sect. \ref{aveDTQW}. In the last section, we calculate the time averaged distribution of Szegedy's walk with related to Ehrenfest model.

%%%%%%%%%%%%%%%%%%%%%%%%%%%%%%%%%%%%%%%%%%%%%%%%%%%%%%%
\section{Definition of the DTQWs}\label{defDTQW}
%%%%%%%%%%%%%%%%%%%%%%%%%%%%%%%%%%%%%%%%%%%%%%%%%%%%%%%
In this paper, we consider DTQWs on the path $P_{n+2}$ with the vertex set $V_{n+2}= \{0,1,\ldots ,n,n+1\}$ and the edge set $E_{n+2}= \{(x,x+1):x=0,1,\ldots ,n\}$. 
%For sake of simplicity of expression, from now on, we use $n$ for suffix of operators although the number of the vertices of $P_{n+2}$ equals $n+2$ because the number of the intermediate vertices in the graph equals $n$. 
In order to define DTQWs, we use a Hilbert space $\mathcal{H}_{n+2}= \mathrm{Span}\{\ket{0,R}, \ket{1,L}, \ket{1,R},\ldots ,\ket{n,L},\ket{n,R},\ket{n+1,L}\}$ with $\ket{x,J}=\ket{x}\otimes \ket{J}\ (x\in V_{n+2}, J\in \{L,R\})$ the tensor product of elements of two orthonormal bases $\{\ket{x}:x\in V_{n+2}\}$ for position of the walker and $\{\ket{L}={}^T [1,0], \ket{R}={}^T [0,1]\}$ for the chirality which means the direction of the motion of the walker where ${}^T \!\!A$ denotes the transpose of a matrix $A$. Then we consider the time evolution operator $U$ on $\mathcal{H}_{n+2}$ defined by $U=SC$ with the coin operator $C$ and the shift operator $S$ (flip-flop type shift) defined as follows:
\begin{align*}
C&=\sum_{x=0}^{n+1}\ket{x}\bra{x}\otimes C_{x},\\
S\ket{x,J}&=
\begin{cases}
\ket{x+1,L}&\text{if}\ \ J=R,\\ 
\ket{x-1,R}&\text{if}\ \ J=L,
\end{cases}
\end{align*}
where $C_{x}\ (x=0,\ldots ,n+1)$ are $2\times 2$ unitary matrices. 

Since the bases of the Hilbert space corresponding to $x=0$ and $x=n+1$ are restricted to $\ket{0,R}$ and $\ket{n+1,L}$, respectively. Thus $C_{0}$ and $C_{n+1}$ are restricted to $C_{0}=c_{0}\ket{R}\bra{R}$ and $C_{n+1}=c_{n+1}\ket{L}\bra{L}$, where $c_{0}, c_{n+1}\in \CM$ with $|c_{0}|=|c_{n+1}|=1$. By the restriction of $C_{0}$ and $C_{n+1}$, the action of the shift operator $S$ is closed on the Hilbert space $\mathcal{H}$. Thus the time evolution operator is the following:
\begin{align*}
U
=
\ket{1}\bra{0}\otimes c_{0}\ket{L}\bra{R}
+
\sum_{x=1}^{n}S\left(\ket{x}\bra{x}\otimes C_{x}\right)
+
\ket{n}\bra{n+1}\otimes c_{n+1}\ket{R}\bra{L}.
\end{align*}
Furthermore if $C_{x} = c_{L}\ket{L}\bra{L} + c_{R}\ket{R}\bra{R}$, where $c_{L}, c_{R}\in \CM$ with $|c_{L}|=|c_{R}|=1$, for some $x=1,\ldots ,n$ then the action of the coin operator $C$ on the Hilbert space $\mathcal{H}_{n+2}$ is separated into that of two parts $\mathrm{Span}\{\ket{0,R}, \ket{1,L}, \ket{1,R},\ldots ,\ket{x-1,L},\ket{x-1,R},\ket{x,L}\}$ and $\mathrm{Span}\{\ket{x,R}, \ket{1,L}, \ket{1,R},\ldots ,\ket{n,L},\ket{n,R},\ket{n+1,L}\}$. So we avoid such a choice of the coin.

Let $X_{t}$ be the position of our quantum walker at time $t$. The probability that the walker with initial state $\ket{\psi}$ is found at time $t$ and the position $x$ is defined by 
\begin{eqnarray*}
\PM_{\ket{\psi}}(X_{t}=x)=\left\lVert \left(\bra{x}\otimes I_{2}\right)U^{t}\ket{\psi}\right\rVert^{2}.
\end{eqnarray*}
In this paper, we consider the DTQW starting from a vertex $0\in V_{n+2}$ and choose the initial chirality state  as $\ket{\psi}_{0}=\ket{0}\otimes \ket{R}$. For the sake of simplicity, we write $\PM_{0}(X_{t}=x)$ for $\PM_{\ket{\psi}_{0}}(X_{t}=x)$. We consider the time averaged distribution 
\begin{eqnarray*}
\bar{p}_{0}(x)=\lim _{T\to \infty }\frac{1}{T}\sum_{t=0}^{T-1}\PM_{0}(X_{t}=x),
\end{eqnarray*} 
where the expectation takes for the choice of the initial chirality state.

%%%%%%%%%%%%%%%%%%%%%%%%%%%%%%%%%%%%%%%%%%%%%%%%%%%%%%%
\section{A connection between DTQWs and DTRWs on the path}\label{connectionQWandRW}
%%%%%%%%%%%%%%%%%%%%%%%%%%%%%%%%%%%%%%%%%%%%%%%%%%%%%%%
In this section, we make a connection between DTQWs and discrete-time random walks (DTRWs) on the path $ P_{n+2}$ using the Jacobi matrices. 

Let $\nu _{1,x}, \nu _{2,x}$ and $\ket{w_{1,x}}, \ket{w_{2,x}}$ be the eigenvalues and the corresponding orthonormal eigenvectors of $C_{x}\ (x=0,\ldots ,n+1)$ which is used in the coin operator $C=\sum_{x=0}^{n+1}\ket{x}\bra{x}\otimes C_{x}$. We consider the spectral decomposition of each unitary matrix $C_{x}$ as follows:
\begin{align}
C_{x}
&=\nu _{1,x}\ket{w_{1,x}}\bra{w_{1,x}}+\nu _{2,x}\ket{w_{2,x}}\bra{w_{2,x}}\notag\\
&=\nu _{1,x}\ket{w_{1,x}}\bra{w_{1,x}}+\nu _{2,x}\left(I_{2}-\ket{w_{1,x}}\bra{w_{1,x}}\right)\notag \\
&=\left(\nu _{1,x}-\nu _{2,x}\right)\ket{w_{1,x}}\bra{w_{1,x}}+\nu _{2,x}I_{2},\label{specC}
\end{align}
where $I_{k}$ is the $k\times k$ identity matrix. Here we use the relation $I_{2}=\ket{w_{1,x}}\bra{w_{1,x}}+\ket{w_{2,x}}\bra{w_{2,x}}$ coming from unitarity of $C_{x}$. This shows that we can represent $C_{x}$ without $\ket{w_{2,x}}$. 

We define the $(n+2)\times (n+2)$ Jacobi matrix $J^{QW}$ for the DTQW as follows:
\begin{align}\label{defJacobiQW}
&(J^{QW})_{x,y}=\overline{(J^{QW})_{y,x}}
\notag
\\
&=
\begin{cases}
\overline{w_{x}(R)}w_{y}(L) & \text{if the pair of vertices $x$ and $y$ $(x<y)$ is adjacent,}\\
0 & \text{otherwise,}
\end{cases}
\end{align}
where $\ket{w_{1,x}}={}^T [w_{x}(L),w_{x}(R)]$ and $\overline{z}$ means the complex conjugate of $z\in \CM$. 
\begin{comment}
In order to connect the DTQW with corresponding DTRW, we should restrict the coin operator as $w_{0}(L)=w_{n+1}(R)=0$. More precisely, we consider the following coin operator:
\begin{align}\label{coinQWonPath}
C_{(n+2)}
&=
\ket{0}\bra{0}\otimes \textrm{diag} (\nu _{2,0}, \nu _{1,0})
+
\sum_{x=1}^{n}\ket{x}\bra{x}\otimes C_{x}
\notag
\\
&+
\ket{n+1}\bra{n+1}\otimes \textrm{diag} (\nu _{1,n+1}, \nu _{2,n+1}).
\end{align}
\end{comment}
In this setting, the corresponding Jacobi matrix is the following:
\begin{comment}
\begin{align}\label{matJacobiQW}
&J_{n+2}^{QW}
\notag
\\
&=
\begin{bmatrix}
0 & \overline{w_{0}(R)}\cdot w_{1}(L) & & & & & \\
w_{0}(R)\cdot \overline{w_{1}(L)} & 0 & \overline{w_{1}(R)}w_{2}(L) & & & \mbox{\smash{\huge\textit{O}}} & \\
 & w_{1}(R)\overline{w_{2}(L)} & \ddots & \ddots & & & \\
 & & \ddots & \ddots & \overline{w_{n-1}(R)}w_{n}(L) & \\
 & & & w_{n-1}(R)\overline{w_{n}(L)} & 0 & \overline{w_{n}(R)}\cdot w_{n+1}(L)\\
\mbox{\smash{\huge\textit{O}}} & & & & w_{n}(R)\cdot \overline{w_{n+1}(L)} & 0
\end{bmatrix}
\end{align}
\end{comment}
\begin{align}\label{matJacobiQW}
&J^{QW}
\notag
\\
&=
\begin{bmatrix}
0 & \overline{w_{0}(R)}\cdot w_{1}(L) & & & \\
w_{0}(R)\cdot \overline{w_{1}(L)} & 0 & \ddots & & \mbox{\smash{\huge\textit{O}}} \\
 & \ddots & \ddots & \ddots & \\
 & & \ddots & 0 & \overline{w_{n}(R)}\cdot w_{n+1}(L)\\
\mbox{\smash{\huge\textit{O}}} & & & w_{n}(R)\cdot \overline{w_{n+1}(L)} & 0
\end{bmatrix}
\end{align}
This Jacobi matrix represents the following inner products:
\begin{align*}
\bra{w_{1,x+1}}U\ket{w_{1,x}}
&=
\nu _{1,x}(J^{QW})_{x+1,x},
\\
\bra{w_{1,x-1}}U\ket{w_{1,x}}
&=
\nu _{1,x}(J^{QW})_{x-1,x}.
\end{align*}
This properties help our spectral analysis in Sect. \ref{spectralDTQW}.

Next, we consider the corresponding DTRW on $P_{n+2}$. Let $p_{x}=|w_{x}(R)|^{2}$ and $q_{x}=|w_{x}(L)|^{2}$ for $x=0,\ldots ,n+1$. We assign $p_{x}$ (resp. $q_{x}$) as the transition probability to the right (resp. left) of the walker on the vertex $x\in V_{n+2}$ in the DTRW. Note that $p_{x}+q_{x}=1$, $p_{x}q_{x}\neq 0$ and the DTRW has reflecting walls at the boundaries $0,n+1\in V_{n+2}$, i.e., the walker moves to the right (left) with probability $1$ at the vertex $0$ ($n+1$), respectively from the definition of the coin operator. This type of DTRW on $P_{n+2}$ is usually called as birth and death chain.

Let $P^{RW}$ be the transition matrix of the DTRW, i.e., $(n+2)\times (n+2)$ matrix with 
\begin{align*}
(P^{RW})_{x,y}=
\begin{cases}
p_{x} &\text{if $y=x+1$},\\
q_{x} &\text{if $y=x-1$},\\
0 &\text{otherwise}.
\end{cases}
\end{align*}
We set an unit vector $\mathbf{\pi }^{1/2} = {}^{T}\left[\pi^{1/2}(0), \ldots ,\pi^{1/2}(n+1)\right]$ such that 
\begin{align*}
\pi^{1/2}(0)=1\times \frac{1}{C_{\pi^{1/2}}}, \quad \pi^{1/2}(x)=\frac{\prod_{y=0}^{x-1}w_{y}(R)}{\prod_{y=1}^{x}w_{y}(L)}\times \frac{1}{C_{\pi^{1/2}}} \quad \text{for $x=1,\ldots ,n+1$,}
\end{align*}
where 
\begin{align*}
C_{\pi^{1/2}} = \sqrt{ 1 + \sum_{x=1}^{n+1}\frac{ \prod_{y=0}^{x-1}p_{y} }{ \prod_{y=1}^{x}q_{y} } }.
\end{align*}
Then we have the following proposition.
\begin{pro}\label{pro:SpecDTRWandDTQW}
$J^{QW}$ and $P^{RW}$ are isospectral. 
\newline
More precisely, if we take $P^{RW}\mathbf{\phi} = \lambda \mathbf{\phi}$ then $J^{QW}\left(D_{\pi^{1/2}}\mathbf{\phi}\right) = \lambda \left(D_{\pi^{1/2}}\mathbf{\phi}\right)$ where $D_{\pi^{1/2}} = \mathrm{diag}\left( \pi^{1/2}(0), \ldots,  \pi^{1/2}(n+1) \right)$. 
\end{pro}
{\bf Proof of Proposition \ref{pro:SpecDTRWandDTQW}.}

We can directly obtain the result. In fact, 
\begin{align*}
\overline{w_{x}(R)}w_{x+1}(L)
&=
p_{x}\times \frac{w_{x+1}(L)}{w_{x}(R)} = p_{x}\times \frac{\pi^{1/2}(x)}{\pi^{1/2}(x+1)},
\\
w_{x}(R)\overline{w_{x+1}(L)}
&=
q_{x+1}\times \frac{w_{x}(R)}{w_{x+1}(L)} = q_{x+1}\times \frac{\pi^{1/2}(x+1)}{\pi^{1/2}(x)}.
\end{align*}
This means that $J^{QW} = D_{\pi^{1/2}}P^{RW}D_{\pi^{1/2}}^{-1}$. From this fact, if we take $P^{RW}\mathbf{\phi} = \lambda \mathbf{\phi}$ then $J^{QW}\left(D_{\pi^{1/2}}\mathbf{\phi}\right) = \lambda \left(D_{\pi^{1/2}}\mathbf{\phi}\right)$.
\qed

\begin{rem}
If we take a vector $\mathbf{\pi } = {}^{T}\left[|\pi^{1/2}(0)|^{2}, \ldots ,|\pi^{1/2}(n+1)|^{2}\right]$ then we have
\begin{align*}
\pi(0)=1\times \frac{1}{C_{\pi}}, \quad \pi(x)=\frac{\prod_{y=0}^{x-1}p_{y}}{\prod_{y=1}^{x}q_{y}}\times \frac{1}{C_{\pi}}\quad \text{for $x=1,\ldots ,n+1$,}
\end{align*}
where 
\begin{align*}
C_{\pi} = 1 + \sum_{x=1}^{n+1}\frac{ \prod_{y=0}^{x-1}p_{y} }{ \prod_{y=1}^{x}q_{y} } .
\end{align*}
This is so-called reversible measure for the DTRW, i.e., it satisfies the following relation:
\begin{align}\label{eq:reversible}
\pi(0)=1,\quad \pi(x) p_{x} = \pi(x+1) q_{x+1}\quad \text{for $x=0,\ldots ,n$.}
\end{align}
\end{rem}

The Jacobi matrix $J^{RW}$ related to the DTRW is defined by
\begin{align}\label{defJacobiRW}
&(J^{RW})_{x,y}=(J^{RW})_{y,x}
\notag
\\
&=
\begin{cases}
\sqrt{p_{x}q_{y}} & \text{if the pair of vertices $x$ and $y$ $(x<y)$ is adjacent,}\\
0 & \text{otherwise.}
\end{cases}
\end{align}
In this case, we have
\begin{align}\label{matJacobiRW}
J^{RW}=
\begin{bmatrix}
0 & \sqrt{1\cdot q_{1}} & & & & & \\
\sqrt{1\cdot q_{1}} & 0 & \sqrt{p_{1}q_{2}} & & & \mbox{\smash{\huge\textit{O}}} & \\
 & \sqrt{p_{1}q_{2}} & \ddots & \ddots & & & \\
 & & \ddots & \ddots & \sqrt{p_{n-1}q_{n}} & \\
 & & & \sqrt{p_{n-1}q_{n}} & 0 & \sqrt{p_{n}\cdot 1}\\
\mbox{\smash{\huge\textit{O}}} & & & & \sqrt{p_{n}\cdot 1} & 0
\end{bmatrix}.
\end{align}
We obtain the following lemma for the two Jacobi matrices $J^{QW}$ and $J^{RW}$:
\begin{lem}\label{lem:JacobiQWandRW}
$J^{QW}$ and $J^{RW}$ are isospectral. In addition, all the eigenvalues are simple. 
\end{lem}
{\bf Proof of Lemma \ref{lem:JacobiQWandRW}.}

As same as the proof of Proposition \ref{pro:SpecDTRWandDTQW}, if we take $D_{\pi} = \mathrm{diag}\left( \pi (0), \ldots,  \pi (n+1) \right)$ then we have $J^{RW} = D_{\pi}^{1/2}P^{RW}D_{\pi}^{-1/2}$. This shows that $J^{RW}$ and $P^{RW}$ are isospectral. Combining with Proposition \ref{pro:SpecDTRWandDTQW}, we obtain the desired result. Simplicity is followed from general argument for the Jacobi matrix (see e.g. Proposition 1.86 of \cite{HoraObata2007}). 
\qed

Combining Proposition \ref{pro:SpecDTRWandDTQW} and Lemma \ref{lem:JacobiQWandRW}, we have a fact that $J^{QW}$ and $P^{RW}$ are isospectral and all the eigenvalues are simple. We also have more detailed information for $\mathrm{Spec}(J^{QW})$ by using that of $\mathrm{Spec}(P^{RW})$. 

\begin{lem}\label{lem:JacobiQWSymSimple}
Every element in $\mathrm{Spec}(J^{QW})\subseteq [-1,1]$ is simple. 
\newline
In addition, if we take $\lambda \in \mathrm{Spec}(J^{QW})$ and the corresponding eigenvector $\mathbf{v}_{\lambda}={}^{T}\left[v_{\lambda}(0) \ldots v_{\lambda}(x) \ldots v_{\lambda}(n+1)\right]$ then $-\lambda \in \mathrm{Spec}(J^{QW})$ and the corresponding eigenvector is $\mathbf{v}_{-\lambda}={}^{T}\left[v_{\lambda}(0) \ldots  (-1)^{x}v_{\lambda}(x) \ldots (-1)^{n+1}v_{\lambda}(n+1)\right]$. Especially, $0\in \mathrm{Spec}(J^{QW})$ if and only if $n$ is odd and $\pm 1\in \mathrm{Spec}(J^{QW})$. 
\end{lem}
{\bf Proof of Lemma \ref{lem:JacobiQWSymSimple}.}

The simplicity is mentioned in Lemma \ref{lem:JacobiQWandRW}. $\mathrm{Spec}(J^{QW})\subseteq [-1,1]$ directly comes from Perron-Frobenius Theorem for $P^{RW}$. If we take $P^{RW}\mathbf{\phi}_{\lambda} = \lambda \mathbf{\phi}_{\lambda}$ with $\mathbf{\phi}_{\lambda}={}^{T}\left[\phi_{\lambda}(0) \ldots \phi_{\lambda}(x) \ldots \phi_{\lambda}(n+1)\right]$ such that $\lambda \neq 0$, we have
\begin{align}\label{eq:EigenJacobi}
q_{x}\phi_{\lambda}(x-1) + p_{x}\phi_{\lambda}(x+1) = \lambda \phi_{\lambda}(x),\quad \text{for $x = 0, \ldots , n+1$},
\end{align}
with a convention $q_{0} = p_{n+1} = 0$. By multiplying $(-1)^{x+1}$ to both side of the equation, we obtain
\begin{align*}
q_{x}(-1)^{x-1}\phi_{\lambda}(x-1) + p_{x}(-1)^{x+1}\phi_{\lambda}(x+1) = -\lambda (-1)^{x}\phi_{\lambda}(x),
\end{align*}
for $x = 0, \ldots , n+1$.

Therefore from Proposition \ref{pro:SpecDTRWandDTQW}, we obtain that if we take $\lambda (\neq 0)\in \mathrm{Spec}(J^{QW})$ and the corresponding eigenvector $\mathbf{v}_{\lambda}={}^{T}\left[v_{\lambda}(0) \ldots v_{\lambda}(x) \ldots v_{\lambda}(n+1)\right]$ then $-\lambda \in \mathrm{Spec}(J^{QW})$ and the corresponding eigenvector is obtained by $\mathbf{v}_{-\lambda}={}^{T}\left[v_{\lambda}(0) \ldots  (-1)^{x}v_{\lambda}(x) \ldots (-1)^{n+1}v_{\lambda}(n+1)\right]$. Thus noting that every element in $\mathrm{Spec}(J^{QW})$ is simple, we have $0\in \mathrm{Spec}(J^{QW})$ if and only if $n$ is odd. Recall that $p_{x}+q_{x}=1$ for all $x=0,\ldots, n+1$, we obtain $\pm 1\in \mathrm{Spec}(J^{QW})$.
\qed

%%%%%%%%%%%%%%%%%%%%%%%%%%%%%%%%%%%%%%%%%%%%%%%%%%%%%%%
\section{A spectral analysis of DTQWs on the path}\label{spectralDTQW}
%%%%%%%%%%%%%%%%%%%%%%%%%%%%%%%%%%%%%%%%%%%%%%%%%%%%%%%
In this section, we give a framework of spectral analysis for DTQWs on $P_{n+2}$. In order to do so, we restrict the coin operator as follows:
\begin{ass}\label{ass:coinQW}
We assume that the coin operator consists of isospectral unitary matrices, i.e., we use 
\begin{align}\label{SpecAnalcoinQWonPath}
C
&=
\nu _{1}\ket{0}\bra{0}\otimes \ket{R}\bra{R}
+
\sum_{x=1}^{n}\ket{x}\bra{x}\otimes \left\{(\nu _{1}-\nu _{2})\ket{w_{x}}\bra{w_{x}}+\nu _{2}I_{2}\right\}
\notag
\\
&+
\nu _{1}\ket{n+1}\bra{n+1}\otimes \ket{L}\bra{L},
\end{align}
as the coin operator, where $\nu_{1}, \nu_{2}\in \CM$ with $\nu_{1}\neq \nu_{2}, |\nu_{1}|=|\nu_{2}|=1$ and each $\ket{w_{x}}={}^T [w_{x}(L),w_{x}(R)]\in \CM^{2}\ (x=1,\ldots ,n+1)$ is a unit vector with $w_{x}(L)w_{x}(R)\neq 0$.
\end{ass}
\begin{comment}
\begin{rem}
The unitarity of the coins $C_{0}$ and $C_{n+1}$ are not necessary. We can use 
\begin{align*}
C_{(n+2)}
&=
\nu _{1}\ket{0}\bra{0}\otimes \ket{R}\bra{R}
+
\sum_{x=1}^{n}\ket{x}\bra{x}\otimes \left\{(\nu _{1}-\nu _{2})\ket{w_{x}}\bra{w_{x}}+\nu _{2}I_{2}\right\}
\notag
\\
&+
\nu _{1}\ket{n+1}\bra{n+1}\otimes \ket{L}\bra{L}
\\
&=
\ket{0}\bra{0}\otimes \textrm{diag} (0, \nu _{1})
+
\sum_{x=1}^{n}\ket{x}\bra{x}\otimes \left\{(\nu _{1}-\nu _{2})\ket{w_{x}}\bra{w_{x}}+\nu _{2}I_{2}\right\}
\notag
\\
&+
\ket{n+1}\bra{n+1}\otimes \textrm{diag} (\nu _{1}, 0)
,
\end{align*}
as the coin operator.
\end{rem}
\end{comment}

Let $\lambda _{m}\ (m=0,\ldots , n+1)$ be the eigenvalues and $\ket{v_{m}}\ (m=0,\ldots , n+1)$ be the corresponding (orthonormal) eigenvectors of $J^{QW}$. For each $\lambda _{m}$ and $\ket{v_{m}}$, we define two vectors
\begin{align*}
\mathbf{a}_{m}
&=
v_{m}(0)\ket{0}\otimes w_{0}(R)\ket{R}
+
\sum_{x=1}^{n}v_{m}(x)\ket{x}\otimes \ket{w_{x}}
\notag
\\
&+
v_{m}(n+1)\ket{n+1}\otimes w_{n+1}(L)\ket{L},
\\
&=
\sum_{x=1}^{n+1}v_{m}(x)w_{x}(L)\ket{x}\otimes \ket{L} + \sum_{x=0}^{n}v_{m}(x)w_{x}(R)\ket{x}\otimes \ket{R},
\\
\mathbf{b}_{m}
&=
S\mathbf{a}_{m}
\\
&=
\sum_{x=1}^{n+1}v_{m}(x-1)w_{x-1}(R)\ket{x}\otimes \ket{L} + \sum_{x=0}^{n}v_{m}(x+1)w_{x+1}(L)\ket{x}\otimes \ket{R},
\end{align*}
where $\ket{v_{m}}={}^{T}\left[v_{m}(0) \ldots v_{m}(n+1)\right]$.
By using $S^{2}=I_{n+2}\otimes I_{2}$, it is easy to see that $C\mathbf{a}_{m}=\nu_{1}\mathbf{a}_{m}$ and then $U\mathbf{a}_{m}=\nu_{1}\mathbf{b}_{m}$. Also we have $C\mathbf{b}_{m}=(\nu_{1}-\nu_{2})\lambda _{m}\mathbf{a}_{m}+\nu_{2}\mathbf{b}_{m}$ and $U\mathbf{b}_{m}=\nu_{2}\mathbf{a}_{m}+(\nu_{1}-\nu_{2})\lambda _{m}\mathbf{b}_{m}$. So we have the following relationship:
\begin{eqnarray}\label{eq:Uab}
U
\begin{bmatrix}
\mathbf{a}_{m}\\
\mathbf{b}_{m}
\end{bmatrix}
=
\begin{bmatrix}
0 & \nu_{1}\\
\nu_{2} & (\nu_{1}-\nu_{2})\lambda _{m}
\end{bmatrix}
\begin{bmatrix}
\mathbf{a}_{m}\\
\mathbf{b}_{m}
\end{bmatrix}.
\end{eqnarray}
We also obtain $|\mathbf{a}_{m}|=|\mathbf{b}_{m}|=1$ and the inner product $(\mathbf{a}_{m}, \mathbf{b}_{m})=\lambda _{m}$. 
This shows that if $\lambda _{m} =\pm 1$ then $\mathbf{b}_{m}=\pm \mathbf{a}_{m}$. Therefore if $\lambda _{m} =\pm 1$ then $U\mathbf{a}_{m}=\pm \nu_{1}\mathbf{a}_{m}$.

For cases with $\lambda_{m}\neq \pm 1$, we see from Eq. (\ref{eq:Uab}) that the operator $U$ is a linear operator acting on the linear space $\text{Span}\ (\mathbf{a}_{m}, \mathbf{b}_{m})$. In order to obtain the eigenvalues and eigenvectors, we take a vector $\alpha \mathbf{a}_{m} + \beta  \mathbf{b}_{m}\in \text{Span}\ (\mathbf{a}_{m}, \mathbf{b}_{m})$. The eigen equation for $U$ is given by $U(\alpha \mathbf{a}_{m} + \beta  \mathbf{b}_{m}) = \mu (\alpha \mathbf{a}_{m} + \beta  \mathbf{b}_{m})$. From Eq. \eqref{eq:Uab}, this is equivalent to 
\begin{eqnarray*}
\begin{bmatrix}
0 & \nu_{2}\\
\nu_{1} & (\nu_{1}-\nu_{2})\lambda_{m}
\end{bmatrix}
\begin{bmatrix}
\alpha \\
\beta
\end{bmatrix}
=
\mu
\begin{bmatrix}
\alpha \\
\beta
\end{bmatrix}
.
\end{eqnarray*}
Therefore we can obtain two eigenvalues $\mu_{\pm m}$ of $U$ which are related to the eigenvalue $\lambda_{m}$ of $J^{QW}$ as solutions of the following quadratic equation:
\begin{eqnarray*}
\mu ^{2}-(\nu_{1}-\nu_{2})\lambda_{m}\mu -\nu_{1}\nu_{2}=0.
\end{eqnarray*}
Also we have the corresponding eigenvectors $\nu_{2}\mathbf{a}_{m}+\mu_{\pm m}\mathbf{b}_{m}$ by setting $\alpha = \nu_{2}, \beta = \mu_{\pm m}$. 

The quadratic equation above is rearranged to
\begin{align*}
\left\{ i \overline{\nu_{1}}^{1/2} \overline{\nu_{2}}^{1/2} \mu \right\}^{2} + 2\Im (\nu_{1}^{1/2} \overline{\nu_{2}}^{1/2})\lambda_{m}\left\{ i \overline{\nu_{1}}^{1/2} \overline{\nu_{2}}^{1/2} \mu \right\} + 1
&=
0.
\end{align*}
Thus we have
\begin{align*}
i \overline{\nu_{1}}^{1/2} \overline{\nu_{2}}^{1/2} \mu _{\pm m}
&=
- \Im (\nu_{1}^{1/2} \overline{\nu_{2}}^{1/2})\lambda_{m} \pm i \sqrt{1-\left( \Im (\nu_{1}^{1/2} \overline{\nu_{2}}^{1/2})\lambda_{m} \right)^{2}}
\\
\mu _{\pm m}
&=
\left( -\nu_{1}\nu_{2} \right)^{1/2}e^{\pm i \theta _{m}},
\end{align*}
where $\cos \theta_{m} = - \Im (\nu_{1}^{1/2} \overline{\nu_{2}}^{1/2})\lambda_{m}$. Therefore if we put $\nu_{j} = e^{i\psi _{j}}$ then the eigenvalues $\mu _{\pm m}$ are given by the following procedure:
\begin{enumerate}
\item
Rescale the eigenvalue $\lambda_{m}$ of $J^{QW}$ as $- \Im (\nu_{1}^{1/2} \overline{\nu_{2}}^{1/2})\lambda_{m} = - \sin [(\psi_{1} - \psi_{2})/2]\times \lambda _{m}$.
\item
Map the rescaled eigenvalue upward and downward to the unit circle on the complex plane.
\item
Take $[(\psi_{1} + \psi_{2} - \pi )/2]$-rotation of the mapped eigenvalues.
\end{enumerate}
If $|- \sin [(\psi_{1} - \psi_{2})/2] | = 1$ then $\psi_{2} = \psi_{1} + 2\pi l + \pi$ for some $l\in \mathbb{Z}$. In this case, $[(\psi_{1} + \psi_{2} - \pi )/2]$-rotation is equal to $[\psi_{1} + \pi l]$-rotation. Combining with Lemma \ref{lem:JacobiQWSymSimple}, we have that every element in $\mathrm{Spec}(U)$ is simple. 

\begin{comment}
For usual Szegedy walk cases, i.e., $\nu _{1} = 1, \nu_{2} = -1$ case, we have $\phi _{1} = 0, \phi_{2} = \pi$. Thus we can omit 1 and 3 of the procedure because $- \sin [(\phi_{1} - \phi_{2})/2] = 1, [(\phi_{1} + \phi_{2} - \pi )/2] = 0$. 
\end{comment}

As a consequence, we obtain the following lemma:
\begin{lem}\label{lem:eigenUformJ}
Every element in $\mathrm{Spec}(U)$ is simple. Let $1=\lambda _{+0} > \lambda _{1} > \cdots > \lambda _{n} > \lambda _{-0} = -1$ be the eigenvalues arranged in decreasing order and $\ket{v_{m}}={}^{T}\left[v_{m}(0) \ldots v_{m}(n+1)\right]\ (m=+0, 1, \ldots , n, -0)$ be the corresponding (orthonormal) eigenvectors of $J^{QW}$. The eigenvalues $\mu_{\pm m}$ and the eigenvectors $\mathbf{u}_{\pm m}\ (m=0, 1, \ldots , n)$ of $U$ are the following:

\begin{enumerate}
\item
$\mu _{\pm 0}=\pm \nu_{1}$ and
\begin{align*}
\mathbf{u}_{\pm 0}
&=
\mathbf{a}_{\pm 0}
\\
&=
\sum_{x=1}^{n+1}v_{\pm 0}(x)w_{x}(L)\ket{x}\otimes \ket{L} + \sum_{x=0}^{n}v_{\pm 0}(x)w_{x}(R)\ket{x}\otimes \ket{R}.
\end{align*}
\item
For $m=1,\ldots , n$, $\mu _{\pm m}=\left( -\nu_{1}\nu_{2} \right)^{1/2}e^{\pm i \theta _{m}}$ where $\cos \theta_{m} = - \Im (\nu_{1}^{1/2} \overline{\nu_{2}}^{1/2})\lambda_{m}$ and 
\begin{align*}
\mathbf{u}_{\pm m}
&=
\nu_{2}\mathbf{a}_{m}+\mu_{\pm m}\mathbf{b}_{m}
\\
&=
\sum_{x=1}^{n+1}\left\{ \nu_{2}v_{m}(x)w_{x}(L)+\mu_{\pm m}v_{m}(x-1)w_{x-1}(R) \right\}\ket{x}\otimes \ket{L}
\\
&+
\sum_{x=0}^{n}\left\{ \nu_{2}v_{m}(x)w_{x}(R)+\mu_{\pm m}v_{m}(x+1)w_{x+1}(L) \right\}\ket{x}\otimes \ket{R}.
\end{align*}
\end{enumerate}
\end{lem} 

\begin{rem}\label{rem:eigenUformJ}
Note that $|\mathbf{a}_{m}|=|\mathbf{b}_{m}|=1$ and $(\mathbf{a}_{m}, \mathbf{b}_{m})=\lambda _{m}\in \mathbb{R}$, we have
\begin{align*}
|\mathbf{u}_{\pm m}|^{2}
&=
|\nu_{2}|^{2}|\mathbf{a}_{m}|^{2} + 2\Re (\overline{\nu_{2}}\mu_{\pm m}(\mathbf{a}_{m},\mathbf{b}_{m})) + |\mu_{\pm m}|^{2}|\mathbf{b}_{m}|^{2}
\\
&=
2\left\{1+\lambda _{m}\Re (\overline{\nu_{2}}\mu_{\pm m})\right\},
\end{align*}
for $m=1,\ldots ,n$.
\end{rem}

%%%%%%%%%%%%%%%%%%%%%%%%%%%%%%%%%%%%%%%%%%%%%%%%%%%%%%%
\section{Time averaged distribution of DTQWs on the path}\label{aveDTQW}
%%%%%%%%%%%%%%%%%%%%%%%%%%%%%%%%%%%%%%%%%%%%%%%%%%%%%%%
Let $\overline{X}_{0}$ be a random variable with distribution $\overline{p}_{0}$, i.e., $\PM (\overline{X}_{0}=x)=\overline{p}_{0}(x)$. Now we estimate the distribution $\overline{p}_{0}$ of the random variable $\overline{X}_{0}$. By the assumption of the choice of the initial state, we have 
\begin{eqnarray*}
\overline{p}_{0}(x)=
\lim _{T\to \infty }\frac{1}{T}\sum_{t=0}^{T-1}
\left\lVert \left(\bra{x}\otimes I_{2}\right)U^{t}(\ket{0}\otimes \ket{R})\right\rVert^{2},
\end{eqnarray*}
Let
\begin{eqnarray*}
\tilde{\mathbf{u}}_{\pm m} =  \frac{ \mathbf{u}_{\pm m} }{ |\mathbf{u}_{\pm m}| } = \sum _{x=0}^{n+1}\ket{x}\otimes \left(u_{x,L}^{(\pm m)}\ket{L}+u_{x,R}^{(\pm m)}\ket{R}\right),
\end{eqnarray*}
be the orthonormal eigenvector corresponding to the eigenvalue $\mu _{\pm m}$ for each $m=0,1,\ldots, n$. 

Using the spectral decomposition $U^{t}=\sum _{m=0}^{n}\sum _{(\pm)}\mu _{\pm m}^{t}\tilde{\mathbf{u}}_{\pm m}\tilde{\mathbf{u}}_{\pm m}^{\dag}$ and $\lim _{T\to \infty }(1/T)\sum_{t=0}^{T-1}e^{i\theta t}=\delta _{0}(\theta )\ (\text{mod} \ 2\pi)$, we obtain 
\begin{eqnarray*}
\overline{p}_{0}(x)=
\sum _{m=0}^{n}\sum _{(\pm)}\left\{(|u_{x,L}^{(\pm m)}|^{2}+|u_{x,R}^{(\pm m)}|^{2})\times |u_{0,R}^{(\pm m)}|^{2}\right\},
\end{eqnarray*}
because all eigenvalues of $U$ are nondegenerate. Using this observation, Lemma \ref{lem:eigenUformJ} and Remark \ref{rem:eigenUformJ}, we build concrete expressions of the components in $\overline{p}_{0}(x)$.

By direct calculation, we have
\begin{align*}
&
\overline{p}_{0}(x)
\\
&=
|v_{+0}(0)|^{2}|v_{+0}(x)|^{2}
+
|v_{-0}(0)|^{2}|v_{-0}(x)|^{2}
\\
&+
\sum_{m=1}^{n} \sum_{(\pm )}
\frac{ \left\{1 + 2\lambda_{m}\Re (\overline{\nu_{2}}\mu_{\pm m}) \right\}|v_{m}(0)|^{2} + q_{1}|v_{m}(1)|^{2} }{ 4\left\{1+\lambda _{m}\Re (\overline{\nu_{2}}\mu_{\pm m})\right\}^{2} }
\\
&\times
\left[ \left\{ 1 + 2\lambda_{m}\Re (\overline{\nu_{2}}\mu_{\pm m})\right\}|v_{m}(x)|^{2} + p_{x-1}|v_{m}(x-1)|^{2} + q_{x+1}|v_{m}(x+1)|^{2} \right],
\end{align*}
with a convention $p_{-1}|v_{m}(-1)|^{2} = q_{n+2}|v_{m}(n+2)|^{2} = 0$. Note that from the derivation procedure of $\mu_{\pm m}$, we have $\Re (\overline{\nu_{2}}\mu_{+m}) = - \Re (\overline{\nu_{2}}\mu_{-(n+1-m)}) $. Combining with Lemma \ref{lem:JacobiQWSymSimple}, we obtain $\lambda_{m}\Re (\overline{\nu_{2}}\mu_{+m}) = \lambda_{n-m}\Re (\overline{\nu_{2}}\mu_{-(n+1-m)}) $. In addition, $q_{1}|v_{m}(1)|^{2} = \lambda_{m}^{2} |v_{m}(0)|^{2}$ from Eq. \eqref{matJacobiQW}. Again from Lemma \ref{lem:JacobiQWSymSimple}, it is observed that $|v_{+0}(x)|^{2} = |v_{-0}(x)|^{2}$. These implies that
\begin{align}\label{eq:AveJacobi}
&
\overline{p}_{0}(x)
\notag
\\
&=
2|v_{+0}(0)|^{2}|v_{+0}(x)|^{2}
\notag
\\
&+
\sum_{m=1}^{n}
\frac{ \left\{1 + 2\lambda_{m}\Re (\overline{\nu_{2}}\mu_{m})  + \lambda _{m}^{2} \right\}|v_{m}(0)|^{2} }{ 2\left\{1+\lambda _{m}\Re (\overline{\nu_{2}}\mu_{m})\right\}^{2} }
\notag
\\
&\times
\big[ \left\{ 1 + 2\lambda_{m}\Re (\overline{\nu_{2}}\mu_{m})  + \lambda _{m}^{2} \right\}|v_{m}(x)|^{2} 
\notag
\\
&+ p_{x-1}|v_{m}(x-1)|^{2} - \lambda _{m}^{2}|v_{m}(x)|^{2} + q_{x+1}|v_{m}(x+1)|^{2} \big].
\end{align}

Recall Proposition \ref{pro:SpecDTRWandDTQW}, if we take $P^{RW}\mathbf{\phi}_{m} = \lambda_{m} \mathbf{\phi}_{m}$ then we have $v_{m}(x) = \pi ^{1/2}(x)\mathbf{\phi}_{m}(x)$. Combining with Eq. \eqref{eq:reversible} and Eq. \eqref{eq:EigenJacobi}, we obtain
\begin{align*}
& 
p_{x-1}|v_{m}(x-1)|^{2} - \lambda _{m}^{2}|v_{m}(x)|^{2} + q_{x+1}|v_{m}(x+1)|^{2} 
\\
%&=
%p_{x-1}\pi (x-1)\mathbf{\phi}_{m}(x-1)^{2} - \lambda _{m}^{2}\pi (x)\mathbf{\phi}_{m}(x)^{2} + q_{x+1}\pi (x+1)\mathbf{\phi}_{m}(x+1)^{2} 
%\\
%&=
%\pi (x) \left\{ q_{x}\mathbf{\phi}_{m}(x-1)^{2} - \lambda _{m}^{2}\mathbf{\phi}_{m}(x)^{2} + p_{x}\mathbf{\phi}_{m}(x+1)^{2} \right\}
%\\
%&=
%\pi (x) \left\{ q_{x}\mathbf{\phi}_{m}(x-1)^{2} - \left[ q_{x}\mathbf{\phi}_{m}(x-1) + p_{x}\mathbf{\phi}_{m}(x+1) \right]^{2} + p_{x}\mathbf{\phi}_{m}(x+1)^{2} \right\}
%\\
%&=
%\pi (x) \left\{ q_{x}(1-q_{x})\mathbf{\phi}_{m}(x-1)^{2} - 2p_{x}q_{x}\mathbf{\phi}_{m}(x-1)\mathbf{\phi}_{m}(x+1) + p_{x}(1-p_{x})\mathbf{\phi}_{m}(x+1)^{2} \right\}
%\\
&=
\pi (x)p_{x}q_{x}\left\{ \mathbf{\phi}_{m}(x-1) - \mathbf{\phi}_{m}(x+1) \right\}^{2}.
\end{align*}
From Eq. \eqref{eq:AveJacobi}, we have the following result:

\begin{thm}\label{thm:Ave}
Let $1=\lambda _{0} > \lambda _{1} > \cdots > \lambda _{n} > \lambda _{n+1} = -1$ be the eigenvalues of $P^{RW}$ arranged in decreasing order and $\mathbf{\phi}_{m}={}^{T}\left[\phi_{m}(0) \ldots \phi_{m}(x) \ldots \phi_{m}(n+1)\right]\ (m=0,\ldots ,n+1)$ be the corresponding eigenvectors with normalization
\begin{align*}
\sum_{x=0}^{n+1}\pi (x)\phi_{m}(x)^{2} = 1\ (m=0,\ldots ,n+1).
\end{align*}
Here 
\begin{align*}
\pi(0)=1\times \frac{1}{C_{\pi}}, \quad \pi(x)=\frac{\prod_{y=0}^{x-1}p_{y}}{\prod_{y=1}^{x}q_{y}}\times \frac{1}{C_{\pi}}\ (x=0,\ldots ,n+1),
\end{align*}
with 
\begin{align*}
C_{\pi} = 1 + \sum_{x=1}^{n+1}\frac{ \prod_{y=0}^{x-1}p_{y} }{ \prod_{y=1}^{x}q_{y} } .
\end{align*}
Under Assumption \ref{ass:coinQW}, the time averaged distribution of DTQW is given by
\begin{align}\label{eq:ThmAve}
&
\overline{p}_{0}(x)
\notag
\\
&=
\pi (0)\pi (x)
\notag
\\
&\times
\Biggr[
2\mathbf{\phi}_{0}(0)^{2}\mathbf{\phi}_{0}(x)^{2}
+
\sum_{m=1}^{n}
\frac{ \left\{1 + 2\lambda_{m}\Re (\overline{\nu_{2}}\mu_{m})  + \lambda _{m}^{2} \right\}\mathbf{\phi}_{m}(0)^{2} }{ 2\left\{1+\lambda _{m}\Re (\overline{\nu_{2}}\mu_{m})\right\}^{2} }
\notag
\\
&\quad \quad \times
\left\{ \left( 1 + 2\lambda_{m}\Re (\overline{\nu_{2}}\mu_{m})  + \lambda _{m}^{2} \right)\mathbf{\phi}_{m}(x)^{2} + p_{x}q_{x}\left\{ \mathbf{\phi}_{m}(x-1) - \mathbf{\phi}_{m}(x+1) \right\}^{2} \right\}
\Biggr].
\end{align}
\end{thm}

As it mentioned before, $\mu _{m}=\left( -\nu_{1}\nu_{2} \right)^{1/2}e^{i \theta _{m}}$ with $\cos \theta_{m} = - \Im (\nu_{1}^{1/2} \overline{\nu_{2}}^{1/2})\lambda_{m}$. If we take $\nu_{2}=-\nu_{1}$ then $\cos \theta_{m} = - \Im (e^{i\pi /2})\lambda _{m} = -\lambda_{m}$. In this case, $\overline{\nu_{2}}\mu _{m}=\left( -\nu_{1}\overline{\nu_{2}} \right)^{1/2}e^{i \theta _{m}} = e^{i \theta _{m}}$. Then we have $\Re (\overline{\nu_{2}}\mu_{m}) = -\lambda_{m}$. Using this fact, we obtain the following result:

\begin{cor}\label{cor:Ave}
If we use the coins with $\nu_{2}=-\nu_{1}$ then
\begin{align}\label{eq:corAve}
&
\overline{p}_{0}(x)
\notag
\\
&=
\pi (0)\pi (x)
\Biggr[
2\mathbf{\phi}_{0}(0)^{2}\mathbf{\phi}_{0}(x)^{2}
\notag
\\
&+
\frac{1}{2}\sum_{m=1}^{n}
\frac{ \mathbf{\phi}_{m}(0)^{2} }{ 1 - \lambda _{m}^{2} }
\left\{ \left( 1 - \lambda _{m}^{2} \right)\mathbf{\phi}_{m}(x)^{2} + p_{x}q_{x}\left\{ \mathbf{\phi}_{m}(x-1) - \mathbf{\phi}_{m}(x+1) \right\}^{2} \right\}
\Biggr].
\end{align}
\end{cor}

\begin{rem}
$\nu_{1}=1, \nu_{2}=-1$ case is referred as the Szegedy's walk.
\end{rem}

\begin{comment}
\begin{rem}
By Lemma \ref{lem:JacobiQWSymSimple} and its proof, Eqs. \eqref{eq:ThmAve} and \eqref{eq:corAve} can express as follows:

For Eq. \eqref{eq:ThmAve},
\begin{align*}
&
\overline{p}_{0}^{(n+2)}(x)
\notag
\\
&=
\pi (0)\pi (x)
\notag
\\
&\quad \quad \quad \times
\Biggr[
2\mathbf{\phi}_{0}(0)^{2}\mathbf{\phi}_{0}(x)^{2}
+
\sum_{m=1}^{\lceil n/2\rceil}
\frac{ \left\{1 + 2\lambda_{m}\Re (\overline{\nu_{2}}\mu_{m})  + \lambda _{m}^{2} \right\}\mathbf{\phi}_{m}(0)^{2} }{ \left\{1+\lambda _{m}\Re (\overline{\nu_{2}}\mu_{m})\right\}^{2} }
\notag
\\
&\quad \quad \quad \quad \quad \times
\left\{ \left( 1 + 2\lambda_{m}\Re (\overline{\nu_{2}}\mu_{m})  + \lambda _{m}^{2} \right)\mathbf{\phi}_{m}(x)^{2} + p_{x}q_{x}\left\{ \mathbf{\phi}_{m}(x-1) - \mathbf{\phi}_{m}(x+1) \right\}^{2} \right\}
\Biggr].
\end{align*}

For Eq. \eqref{eq:corAve},
\begin{align*}
&
\overline{p}_{0}^{(n+2)}(x)
\notag
\\
&=
\pi (0)\pi (x)
\Biggr[
2\mathbf{\phi}_{0}(0)^{2}\mathbf{\phi}_{0}(x)^{2}
+
\sum_{m=1}^{\lceil n/2\rceil}
\frac{ \mathbf{\phi}_{m}(0)^{2} }{ 1 - \lambda _{m}^{2} }
\left\{ \left( 1 - \lambda _{m}^{2} \right)\mathbf{\phi}_{m}(x)^{2} + p_{x}q_{x}\left\{ \mathbf{\phi}_{m}(x-1) - \mathbf{\phi}_{m}(x+1) \right\}^{2} \right\}
\Biggr].
\end{align*}
Here $\lceil r \rceil$ means the smallest integer which is greater than or equal to $r$. 
\end{rem}
\end{comment}

The proof of Theorem \ref{thm:Ave} is to calculate the concrete forms of $|u_{x,L}^{(\pm m)}|^{2}+|u_{x,R}^{(\pm m)}|^{2}$ for $m=0,1, \ldots , n$. These are nothing but stationary distributions for the DTQW. Combining with Lemma \ref{lem:JacobiQWSymSimple} and its proof, there are at least $\lceil n/2 \rceil +1$ numbers of stationary distributions for the DTQW, where $\lceil r \rceil$ means the smallest integer which is greater than or equal to $r$.
\begin{pro}\label{pro:StationaryMeas}
There are at least $\lceil n/2 \rceil +1$ numbers of stationary distributions $\mathfrak{p}_{m}\ (m=0,1,\ldots \lceil n/2 \rceil)$ for the DTQW as follows:
\begin{align*}
\mathfrak{p}_{0}(x)
&= \pi (x)\phi_{0}(x)^{2},
\\
\mathfrak{p}_{m}(x)
&= \pi (x)\frac{ \left( 1 + 2\lambda_{m}\Re (\overline{\nu_{2}}\mu_{m})  + \lambda _{m}^{2} \right)\mathbf{\phi}_{m}(x)^{2} + p_{x}q_{x}\left\{ \mathbf{\phi}_{m}(x-1) - \mathbf{\phi}_{m}(x+1) \right\}^{2} }{ 4\left\{1+\lambda _{m}\Re (\overline{\nu_{2}}\mu_{m})\right\}^{2} }
\notag
\\
&(m=1,\ldots ,\lceil n/2 \rceil).
\end{align*}
In particular, for $\nu_{2}=-\nu_{1}$ case, 
\begin{align*}
\mathfrak{p}_{0}(x)
&= \pi (x)\phi_{0}(x)^{2},
\\
\mathfrak{p}_{m}(x)
&= \pi (x)\frac{ \left( 1 - \lambda _{m}^{2} \right)\mathbf{\phi}_{m}(x)^{2} + p_{x}q_{x}\left\{ \mathbf{\phi}_{m}(x-1) - \mathbf{\phi}_{m}(x+1) \right\}^{2} }{ 4(1-\lambda _{m}^{2}) }
\notag
\\
&(m=1,\ldots ,\lceil n/2 \rceil).
\end{align*}
\end{pro}

\section{Szegedy's walk related to the Ehrenfest model}
In this section, we consider the Szegedy's walk related to the Ehrenfest model \cite{EhrenfestEhrenfest1907} which is defined by $p_{x}=1-x/(n+1), q_{x}=x/(n+1)$. This corresponds directly to the simple random walk on hypercube \cite{DiaconisShahshahani1987}. There is a result on time averaged distribution for this model \cite{MarquezinoEtAl2008}. In this section, we give more concrete form of the time averaged distribution. 

For Ehrenfest model, it is known that
\begin{align*}
\lambda_{m}
&=
1 - \frac{2m}{n+1}, 
\\
\mathbf{\phi}_{m}(x)
&=
\binom{n+1}{m}^{-1/2}\sum_{j=0}^{m}(-1)^{j}\binom{n+1-x}{m-j}\binom{x}{j},
\\
\pi (x)
&=
\binom{n+1}{x}2^{-(n+1)},
\end{align*}
with the conventions
\begin{align*}
\binom{a}{b} = 0,\quad \text{if } a<b,\quad \binom{a}{0} = 1.
\end{align*}
Note that $\mathbf{\phi}_{m}(x)$ is referred as Krawtchouk polynomial \cite{Feinsilver2016, FeinsilverFitzgerald1996}. %What we want to know is that whether $\overline{p}_{0}^{(n+2)}(x)/\binom{n+1}{x}$ defined by Eq. \eqref{eq:corAve} equals Eq. (13) with $n \leftrightarrow n+1$ in \cite{MarquezinoEtAl2008} or not. We have already checked that it equals Eq. (14).  

Note that $\mathbf{\phi}_{m}(0)=\binom{n+1}{m}^{1/2}$. Thus $\mathbf{\phi}_{m}(0)\mathbf{\phi}_{m}(x)= \sum_{j=0}^{m}(-1)^{j}\binom{n+1-x}{m-j}\binom{x}{j}$. Let $N=n+1$ and 
\begin{align*}
\mathbf{\phi}_{m}^{(N)}(x)
&=
\sum_{j=0}^{m}(-1)^{j}\binom{N-x}{m-j}\binom{x}{j}.
\end{align*}
By the binomial relation, we obtain the following identities:
\begin{align*}
\mathbf{\phi}_{m-1}^{(N)}(x) + \mathbf{\phi}_{m}^{(N)}(x)
&= 
\mathbf{\phi}_{m}^{(N+1)}(x),
\\
\mathbf{\phi}_{m}^{(N)}(x) - \mathbf{\phi}_{m-1}^{(N)}(x)
&= 
\mathbf{\phi}_{m}^{(N+1)}(x+1).
\end{align*}
By adding and subtracting above equations we obtain
\begin{align*}
\mathbf{\phi}_{m}^{(N)}(x) + \mathbf{\phi}_{m}^{(N)}(x+1)
&= 
2\mathbf{\phi}_{m}^{(N-1)}(x),
\\
\mathbf{\phi}_{m}^{(N)}(x) - \mathbf{\phi}_{m}^{(N)}(x+1)
&= 
2\mathbf{\phi}_{m-1}^{(N-1)}(x).
\end{align*}
Using these identities, we have 
\begin{align*}
&\mathbf{\phi}_{m}^{(N)}(x-1) - \mathbf{\phi}_{m}^{(N)}(x+1)
\\
&=
\left( 2\mathbf{\phi}_{m}^{(N-1)}(x-1) -  \mathbf{\phi}_{m}^{(N)}(x) \right) + \left( 2\mathbf{\phi}_{m-1}^{(N-1)}(x) - \mathbf{\phi}_{m}^{(N)}(x) \right)
\\
&=
2\left\{ \mathbf{\phi}_{m}^{(N-1)}(x-1) + \mathbf{\phi}_{m-1}^{(N-1)}(x) - \mathbf{\phi}_{m}^{(N)}(x) \right\}
\\
&=
2\left\{ \mathbf{\phi}_{m}^{(N-1)}(x-1) - \mathbf{\phi}_{m}^{(N-1)}(x) \right\}
\\
&=
4\mathbf{\phi}_{m-1}^{(N-2)}(x-1).
\end{align*}
%Note that $1-\lambda_{m}^{2} = 4p_{m}q_{m}$, we obtain 
Thus we obtain
\begin{align*}
& 
\frac{ \mathbf{\phi}_{m}(0)^{2} }{ 1 - \lambda _{m}^{2} }
\left\{ \left( 1 - \lambda _{m}^{2} \right)\mathbf{\phi}_{m}(x)^{2} + p_{x}q_{x}\left\{ \mathbf{\phi}_{m}(x-1) - \mathbf{\phi}_{m}(x+1) \right\}^{2} \right\}
\\
&=
\left( \mathbf{\phi}_{m}^{(N)}(x) \right)^{2} + \frac{ p_{x}q_{x} }{ 1 - \lambda _{m}^{2} }\left( 4\mathbf{\phi}_{m-1}^{(N-2)}(x-1) \right)^{2}.
\end{align*}

It is known \cite{Feinsilver2016, FeinsilverFitzgerald1996} that
\begin{align*} 
\sum _{m=0}^{N}\left( \mathbf{\phi}_{m}^{(N)}(x) \right)^{2} = \frac{ \binom{2N-2x}{N-x}\binom{2x}{x} }{ \binom{N}{x} }. 
\end{align*}
In this case, we obtain
\begin{align}\label{eq:aveEhrenfest}
&\overline{p}_{0}(x)
\notag
\\
&=
\frac{ \binom{N}{x} }{ 2^{2N} } \left[ 1 + \frac{ 1 }{ 2\binom{N}{x} }\binom{2N-2x}{N-x}\binom{2x}{x} + \frac{1}{2}\sum_{m=1}^{N-1}\frac{ p_{x}q_{x} }{ 1 - \lambda _{m}^{2} }\left( 4\mathbf{\phi}_{m-1}^{(N-2)}(x-1) \right)^{2} \right] 
.
\end{align}
The second term is nothing but the discrete version of arcsine law with a coefficient $1/2$. Note that 
\begin{align*}
\frac{ \Gamma (N+1/2) }{2\sqrt{\pi }N\Gamma(N)} &= \frac{(2N-1)!!}{ 2^{N+1}N! } = \frac{(2N-1)!}{ 2^{2N}(N-1)!N! } = \frac{(2N)!}{ 2^{2N+1}N!N! } =  \frac{1}{ 2^{2N+1} }\binom{2N}{N}
\notag
\\
&= \overline{p}_{0}(0) - \frac{ 1 }{ 2^{2N} }.
\end{align*}
This shows that this result is consistent with Eq. (14) in \cite{MarquezinoEtAl2008}.

\section{Conclusion}
In this paper, we considered DTQWs on the path graph. We gave a procedure for construction of the corresponding birth and death chain at first. Using this correspondence, we obtained a formal form for the time averaged distribution of the DTQWs with isospectral but need not be the same coins. This form consists of the pair of eigenvalues of the coins as well as the eigenvalues, eigenvectors and the stationary distribution of the corresponding birth and death chain. Therefore we will be able to compare various DTQWs and the corresponding birth and death chains directly. One of an interesting example of the corresponding birth and death chain is the Ehrenfest model. We have shown that the time averaged distribution of the Szegedy's walk corresponding to the Ehrenfest model includes the discrete version of arcsine law. Making the scaling limit of this model clear can be an interesting future problem. 

\par
\
\par\noindent
{\bf Acknowledgments.} 

We thank the anonymous referees for their careful reading of our manuscript and their fruitful comments and suggestions. C. L. H. was supported in part by the Ministry of Science and Technology (MoST) of the Republic of China under Grants MoST 105-2112-M-032-003. Y. I. was supported by the Grant-in-Aid for Young Scientists (B) of Japan Society for the Promotion of Science (Grant No. 16K17652). N. K. was supported by the Grant-in-Aid for Challenging Exploratory Research of Japan Society for the Promotion of Science (Grant No. 15K13443). E. S. was supported by the Grant-in-Aid for Young Scientists (B) and the Grant-in-Aid for Scientific Research (B) of Japan Society for the Promotion of Science (Grant No. 16K17637, 16H03939).
%\par
%\
%\par
\begin{small}

\end{small}


\begin{thebibliography}{000}


\bibitem{AharonovEtAl2001}
Aharonov, D., Ambainis, A., Kempe, J., Vazirani, U. V.: 
Quantum walks on graphs.
Proc. of the 33rd Annual ACM Symposium on Theory of Computing (STOC '01), 50--59 (2001).


\bibitem{AraiEtAl2016}
Arai, T., Ho, C.-L., Ide, Y., Konno, N.: 
Periodicity for space-inhomogeneous quantum walks on the cycle.
Yokohama Math. J. {\bf 62}, 39--50 (2016).


\bibitem{BaluLiuVAndraca2017}
Balu, R., Liu, C., Venegas-Andraca, S.:
Probability distributions for Markov chains based quantum walks.
J. Phys. A: Math. Theor. {\bf 51}, 035301 (2008).


\bibitem{BednarskaEtAl2003}
Bednarska, M., Grudka, A., Kurzy\'nski, P., \L uczak, T., W\'ojcik, A.: 
Quantum walks on cycles.
Phys. Lett. A {\bf 317}, 21--25 (2003).


\bibitem{DiaconisShahshahani1987}
Diaconis, P., Shahshahani, M.:
Time to reach stationary in the Bernoulli-Laplace diffusion model.
SIAM J. Math. Anal. {\bf 18}, 208--218 (1987).

\bibitem{EhrenfestEhrenfest1907}
Ehrenfest, P., Ehrenfest, T.:
\"Uber zwei bekannte Einw\"ande gegen das Boltzmannsche H-Theorem.
Phys. Zeits. {\bf 8}, 311--314 (1907).

\bibitem{Feinsilver2016}
Feinsilver, P.:
Sums of squares of Krawtchouk polynomials, Catalan numbers, and some algebras over the boolean lattice. 
arXiv:1603.07023v1 (2016).

\bibitem{FeinsilverFitzgerald1996}
Feinsilver, P., Fitzgerald, R.:
The spectrum of symmetric Krawtchouk matrices.
Lin. Alg. \& Appl. {\bf 235}, 121--139 (1996).


\bibitem{HoraObata2007}
Hora, A., Obata, N., 
Quantum Probability and Spectral Analysis of Graphs. 
Springer (2007).


\bibitem{IdeKonnoSegawa2012}
Ide, Y., Konno, N., Segawa, E.: 
Time averaged distribution of a discrete-time quantum walk on the path.
Quantum Inf. Process. {\bf 11} (5), 1207--1218 (2012).

\bibitem{IdeEtAl2014}
Ide, Y., Konno, N., Segawa, E., Xu, X.-P.: 
Localization of discrete time quantum walks on the glued trees.
Entropy {\bf 16} (3), 1501--1514 (2014).


\bibitem{Kempe2003} 
Kempe, J.: 
Quantum random walks - an introductory overview. Contemporary Physics {\bf 44},  307--327 (2003).


\bibitem{Kendon2007} 
Kendon, V.: 
Decoherence in quantum walks - a review. Math. Struct. in Comp. Sci. {\bf 17}, 1169--1220 (2007).


\bibitem{Konno2008b} 
Konno, N.: 
Quantum Walks. In: Quantum Potential Theory, Franz, U., and Sch\"urmann, M., Eds., Lecture Notes in Mathematics: Vol. 1954, pp. 309--452, Springer-Verlag, Heidelberg (2008).


\bibitem{ManouchehriWang2013} 
Manouchehri, K., Wang, J.: 
Physical Implementation of Quantum Walks, Springer (2013).


\bibitem{MarquezinoEtAl2008}
Marquezino, F. L., Portugal, R., Abal, G., Donangelo, R.:
Mixing times in quantum walks on the hypercube. 
Phys. Rev. A {\bf 77}, 042312 (2008).


\bibitem{Portugal2013} 
Portugal, R.: 
Quantum Walks and Search Algorithms, Springer (2013).


\bibitem{PortugalSegawa2017}
Portugal, R., Segawa, E.:
Coined Quantum Walks as Quantum Markov Chains.
Interdisciplinary Information Sciences {\bf 23} (1), 119--125 (2017).


\bibitem{Segawa2013} 
Segawa, E.: 
Localization of quantum walks induced by recurrence properties of random walks. 
J. Comput. Nanosci. {\bf 10}, 1583--1590 (2013).


\bibitem{Szegedy2004} 
Szegedy, M.: 
Quantum speed-up of Markov chain based algorithms. 
Proc. of the 45th Annual IEEE Symposium on Foundations of Computer Science (FOCS '04), 32--41 (2004).


\bibitem{VAndraca2008} 
Venegas-Andraca, S. E.: 
Quantum Walks for Computer Scientists, Morgan and Claypool (2008).


\bibitem{VAndraca2012} 
Venegas-Andraca, S. E.: 
Quantum walks: a comprehensive review, Quantum Inf. Process. {\bf 11}, 1015--1106 (2012).


\end{thebibliography}
\end{document}